\documentclass[journal=jacsat,manuscript=article]{achemso}

\usepackage[version=3]{mhchem} 
\usepackage{xcolor}
\usepackage{booktabs}
\usepackage[numbers]{natbib}  
\usepackage{xcolor}            
\usepackage{graphicx}          
\usepackage{float}
\usepackage{subcaption}

\usepackage{graphicx,lipsum}  
\usepackage{subcaption,xcolor}  

\author{Sukanya Jana}
\affiliation{School of Chemical Sciences, Indian Association for the Cultivation of Science, Kolkata-700032, India}

\author{Santu Nath}
\affiliation{Tata Institute of Fundamental Research, Hyderabad-500046, India}

\author{Pintu Mandal}
\email{pintuphys@gmail.com}
\affiliation{Department of Physics, St. Paul's Cathedral Mission College, Kolkata-700009, India}

\author{Nabanita Deb}
\email{nabanita.deb@iacs.res.in}
\affiliation{School of Chemical Sciences, Indian Association for the Cultivation of Science, Kolkata-700032, India}

\title[An \textsf{achemso} demo]
 {
 Effect of Rod Asymmetry on Field Distortions and Transmission Characteristics in Linear Quadrupole Mass Filters
 }
\abbreviations{IR,NMR,UV}

\begin{document}

\begin{abstract}
  This article reports the appearance of the octupole and hexadecapole components in the radial potential, which are otherwise absent in a linear quadrupole mass filter of symmetric round-rod configuration, if a small structural asymmetry is introduced in one electrode. Using the commercial ion simulation package SIMION, it is shown that the weightage of the octupole and hexadecapole potential increases with an increase in the asymmetry factor. A consistent correlation is observed between the octupole and dodecapole components, and between the hexadecapole and icosapole components with asymmetry added to different rod-to-field radius ratios. The transmission characteristics of the mass filter are studied for different asymmetry factors and rod-to-field radius ratios. The variation in the resolution parameter, as derived from the transmission contour, exhibits a strong correlation with the combined weightage of the multipole fields.      
\end{abstract}

\section{Introduction}

Since its introduction in 1953 by Paul et al.~\cite{paul1953neues,paul1955elektrische}, the quadrupole mass filter (QMF) has undergone significant advancements \cite{douglas2009,richards1973new,march1989quadrupole,paul1953neues}. While initial studies emphasized ideal hyperbolic electrodes, practical considerations favored circular rods due to fabrication constraints~\cite{dawson2013quadrupole,denison1971operating,gibson2001numerical}. QMF with circular rods appears with dodecapole and icosapole components as the dominating higher order potential terms which distort the desired quadrupole potential. The octupole and hexadecapole components are usually absent in symmetric circular rod configuration.~\cite{dawson2013quadrupole} Significant efforts were directed toward optimizing the rod-to-field radius ratio ($\eta$) to suppress higher-order field components, with critical contributions from Dayton, Denison, Lee-Whiting, Yamazaki, and Reuben~\cite{dayton1954measurement, denison1971operating, lee1971semi, reuben1994exact}. Several studies investigated conditions minimizing the dodecapole component, further explored by March and Stafford \cite{march2000quadrupole, stafford1984recent}. Douglas and Konenkov reported the optimal performance at $\eta = 1.13$, revising the previously accepted value of $\eta = 1.14511$\cite{douglas2002influence, dayton1954measurement,denison1971operating,reuben1996ion,lee1971semi}. Konenkov proposed incorporating hexapole fields into QMF designs to enhance resolution~\cite{konenkov2006linear}. 

More recently, approaches to mitigate multipole distortions have been developed, including the harmonic balancing strategy by Konenkov et al. targeting dodecapole and icosapole terms~\cite{sysoev2022}, and Taylor and Gibson's demonstration of improved performance through modified electrical configurations~\cite{taylor2008prediction}. Mechanical asymmetries, such as inward displacement of a single rod, can generate spurious precursor peaks, which may be mitigated through voltage compensation~\cite{taylor2008prediction}. Ding et al. introduced a controlled octupole component to improve resolution~\cite{ding2003quadrupole}, while Bugrov investigated the role of octupole fields under a negative DC bias, showing notable resolution enhancements~\cite{bugrov2023simulation}. A recent experimental study reports octupole nonlinear resonance in a linear ion trap with one asymmetric electrode~\cite{mandal2024}. The present work focuses on a systematic study of the multipole fields, their interplay and transmission characteristics in similarly asymmetric QMFs, an area that has not yet been explored.

\begin{figure}
    \centering
    \includegraphics[width=1\linewidth]{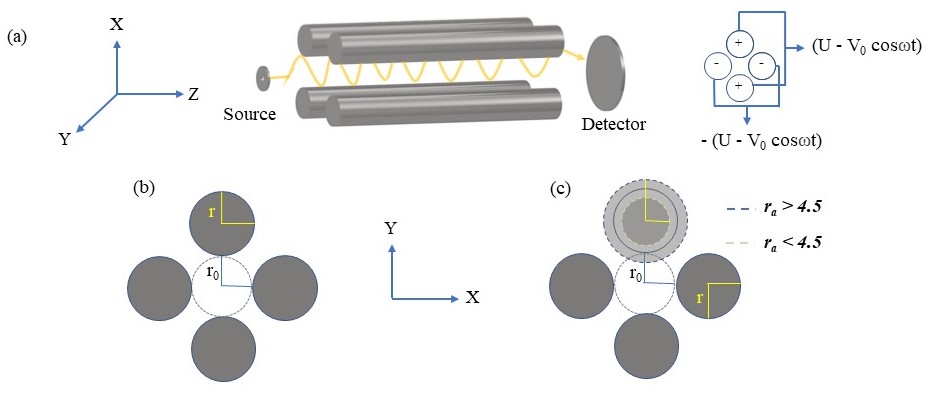}
    \caption{(a) Schematic of the quadrupole mass filter setup. Electrode arrangements in the $xy$ plane in (b) symmetric and (c) asymmetric setup. $r$ and $r_a$ represent the radius of the symmetric and asymmetric electrode respectively.}
    \label{fig1}
\end{figure}

The general form of the electric potential inside the linear QMF is~\cite{dawson2013quadrupole,ding2003quadrupole}:
\begin{equation}
    \Phi(x,y,t)=\sum_{N=0}^{\infty}\phi_{N}(x,y)(U-V_{0}\cos\Omega t),
    \label{eq1}
\end{equation}
where
\begin{equation}
    \phi_{N}(x,y)=Re\ A_{N}\left(\frac{z}{r_{0}}\right)^{N}, \,z=x+iy.
     \label{eq2}
\end{equation}
$A_{N}$ is the coefficient of $N^{th}$ order multipole and $r_0$ is the radius of the inscribed circle as shown in fig. \ref{fig1}. For a pure quadrupole field $N=2$, $A_2=1$ and hence, $\phi_{2}(x,y)=(x^2-y^2)/r_{0}^{2}$. The corresponding equations of motion of an ion of charge $e$ and mass $m$ in the radial plane are described by the Matheiu differential equation as
\begin{equation}
  \frac{d^2 u}{d\xi^2} + (a_u - 2q_u \cos 2\xi)u = 0,
   \label{eq3}
\end{equation}
where \(u=x,y\) and 
\begin{equation}
    a_x = -a_y = \frac{8eU}{m r_0^2 \Omega^2}, \quad q_x = -q_y = \frac{4eV}{m r_0^2 \Omega^2}, \quad \xi=\Omega t/2.
     \label{eq4}
\end{equation}

A rigorous analysis of field distortions and their effects on ion dynamics requires an explicit formulation of the time-dependent electric potential within the QMF. This formulation enables decomposition into multipole components, allowing quantitative assessment of deviations from the ideal quadrupolar field. 

Ion trajectory simulations typically employ either direct numerical integration of the equations of motion or matrix-based stability analysis~\cite{douglas2009, konenkov2002matrix, titov1998detailed, dawson2013quadrupole}. Harmonic amplitudes ($A_N$) are extracted using established numerical techniques~\cite{douglas1999spatial, konenkov2010spatial}. Alternatively, commercial ion simulation package SIMION can also be used to simulate the potential surface in linear quadrupole mass filter. It uses a potential array approach, where potentials are arranged as mesh points or grids. Electrode boundaries are defined by assigning fixed potentials to selected array points, and higher points densities are taken near the boundaries to improve simulation accuracy. The potential at other grid points are computed by solving Laplace's equation using the finite difference method.~\cite{brkic2009development,blaum1998properties,hu2024simulation,shinholt2014frequency,yang2023simulation} 

In this work, the effect of asymmetry or imperfection in the QMF design on the higher order multipole field contribution and consequently its transmission characteristic is studied using SIMION. The simulated potential surface is fitted to eq.~\ref{eq2} and the coefficients of different order terms have been determined thereby. The results thus obtained are corroborated with the results reported earlier for some symmetric setups corresponding to $\eta=1.12, 1.13, 1.14$ and $1.15$.~\cite{douglas2002influence} A similar approach is followed to determine the weightage of the multipole terms present in the QMF with one asymmetric electrode. The octupole ($N=4$) and hexadecapole ($N=8$) potential terms are found to appear in such asymmetric setups, and the corresponding coefficients $A_4$ and $A_8$ increase with increasing the asymmetry parameter. The transmission characteristics of symmetric and asymmetric QMFs are reported in the following sections for various geometric factors, stability parameters, driving frequency, and asymmetric designs.

\section{Symmetric QMF}
Fig.~\ref{fig1}(a) depicts schematic of conventional linear QMF with symmetric circular rod structure. Four identical electrodes are placed symmetrically on $xy$ plane as shown in fig.~\ref{fig1}(b). The key geometric parameter is defined as $\eta=r/r_0$, the ratio of the electrode radius $r$ and the inscribed circle radius or the field radius $r_0$ which is set as $5.0$~mm in our work. The setup is housed inside a grounded chamber of radius $4r_0=20$~mm.

For potential surface simulations, a DC potential of $+1$~V is applied to one diagonal pair of rods and $-1$~V to the opposite pair. All transmission studies presented here employ an ion beam with kinetic energy of $0.5$~eV and a mass-to-charge ratio of $m/e = 40$~u/C. The ions are initially distributed uniformly within a $0.5$~mm radius in the $xy$-plane, with time of birth randomly selected in the interval $0$ to $0.5~\mu$s. A fixed radiofrequency (rf) drive of $\omega/2\pi = 2.0$~MHz is used unless stated otherwise, ensuring that each ion experiences over 100 rf cycles while traversing the QMF of length $80$~mm.

\subsection{Multipole fields}

The potential surface within the symmetric QMF was simulated using SIMION for various geometric configurations with $\eta = 1.12$, $1.13$, $1.14$, and $1.15$, based on the setup described above. The resulting potential surfaces were fitted to eq.~\ref{eq2}, and the multipole coefficients $A_N$ were extracted for $N$ up to 10. Coefficients smaller than the lowest reported value in table \ref{table1} are considered negligible and set to zero. As expected, only the dodecapole ($N=6$) and icosapole ($N=10$) terms appear in addition to the dominant quadrupole component, consistent with the symmetry of the circular rod QMF geometry~\cite{dawson2013quadrupole}. 

The extracted coefficients show excellent agreement with previously reported values~\cite{douglas2002influence}. Our study, based on the graph of $A_N$ versus $\eta$, reveals that $A_6$ becomes zero at $\eta=1.144$ as compared to previous reports at $\eta=1.14511$ \cite{douglas2002influence,sysoev2022}. These findings underscore the precision and robustness of our approach, which accounts for variations in computational methods, boundary conditions, and numerical approximations. Such insights contribute to a deeper and easier understanding of QMF geometries and further refine the established theoretical predictions.

\begin{table*}[h!]
\centering
\caption{\label{tab:A-n}Amplitude $A_N$ vs. $r/r_0$ for a symmetrical arrangement of identical electrodes}
\begin{tabular}{c|cc|cc|cc}
\hline
$r/r_0$ & \multicolumn{2}{c|}{$A_2$} & \multicolumn{2}{c|}{$A_6$$\times10^{4}$} & \multicolumn{2}{c}{$A_{10}$$\times10^{3}$} \\ 
\cline{2-7}
        & this work  & ref [\cite{douglas2002influence}] & this work  & ref [\cite{douglas2002influence}] & this work  & ref[\cite{douglas2002influence}] \\ 
\hline

1.12    & 1.00085  & 1.00108        & 16.21 & 16.71 & -2.420 & -2.420 \\
1.13    & 1.00152  & 1.00176        & 9.51  & 10.00  & -2.440 & -2.440 \\
1.14    & 1.00219  & 1.00242        & 2.95 & 3.36  & -2.443 & -2.445 \\
1.15    & 1.00284  & 1.00307        & -3.61 & -3.19 & -2.455 & -2.453 \\
\hline
\end{tabular}
\label{table1}
\end{table*}

\subsection{QMF transmission}

QMF transmission is defined as the ratio of the number of ions that reach the detector (or exit plane of the QMF) without hitting the electrodes to the total number of ions initially injected into the system, under given operating conditions such as the scan parameter, $\lambda = a/2q$, the driving frequency, $\omega$ and the geometric factor, $\eta$.  The influence of these operational parameters is examined to identify conditions that maximize the transmission efficiency. A series of SIMION simulations were performed for the optimized geometry at $\eta = 1.14$, which exhibits minimal field distortion.

\begin{figure}
    \centering
    \begin{subfigure}[b]{0.48\textwidth}
        \begin{picture}(0,0)
            \put(-10,180){\textbf{(a)}}
        \end{picture}
        \includegraphics[width=\textwidth]{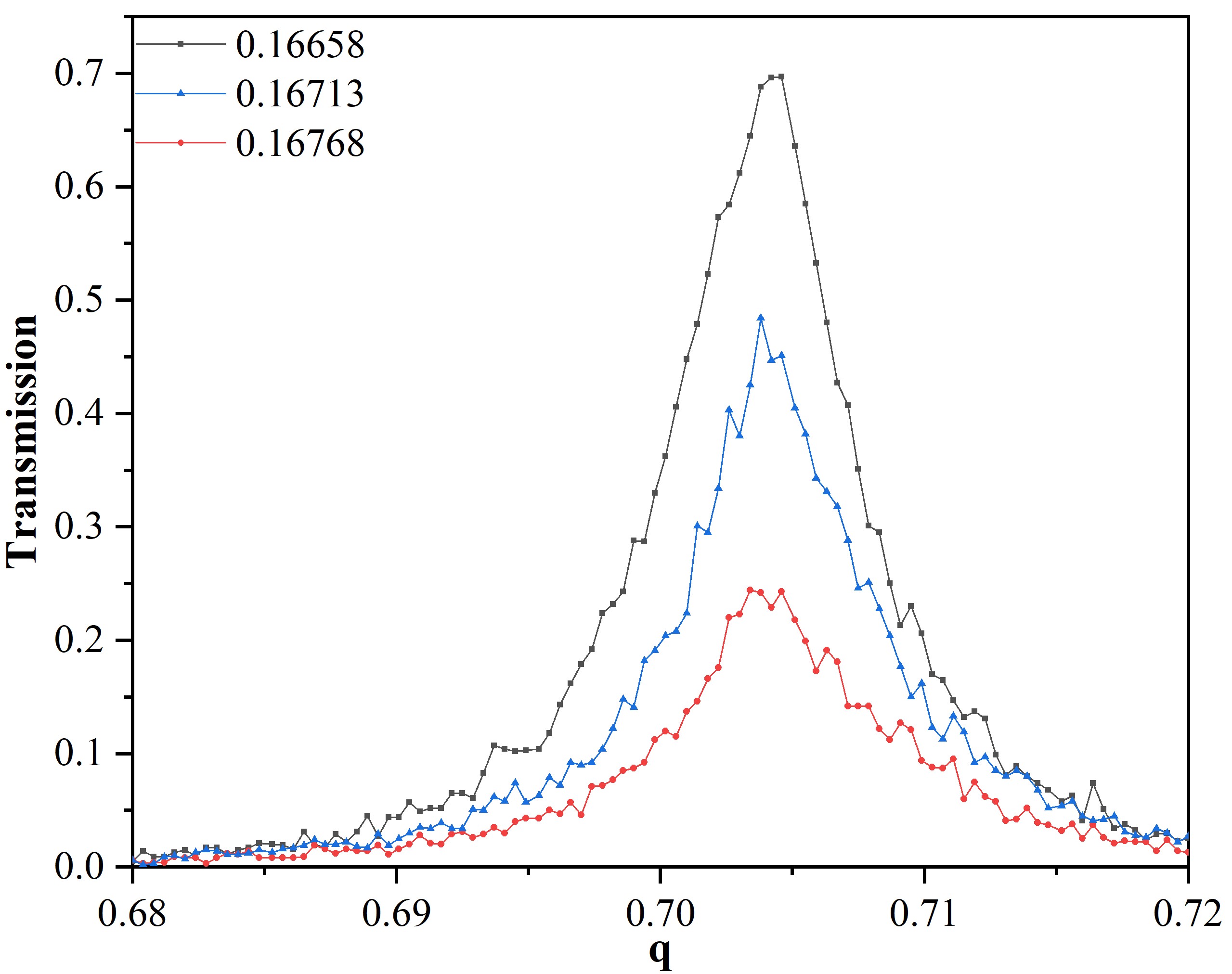}
        \label{fig:a}
    \end{subfigure}
    \hspace{2mm} 
    \begin{subfigure}[b]{0.48\textwidth}
        \begin{picture}(0,0)
            \put(-10,185){\textbf{(b)}}
        \end{picture}
        \includegraphics[width=\textwidth]{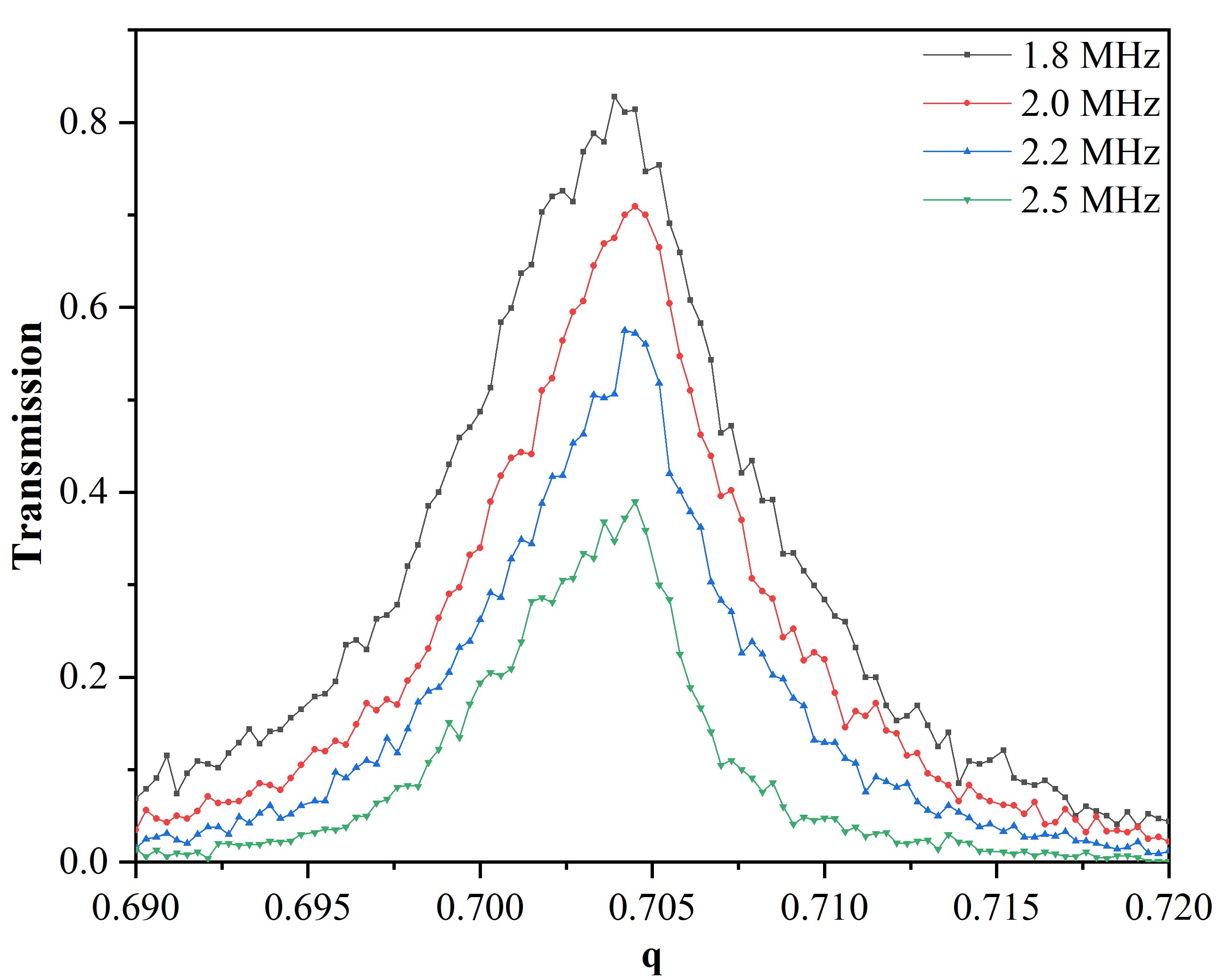}
        \label{fig:b}
    \end{subfigure}
    \caption{Transmission for $\eta=1.14$ at (a) different scan parameter $\lambda$  and $\omega/2\pi=2$~MHz. (b) different driving frequency and $\lambda=0.16658$.}
    \label{fig2}
\end{figure}
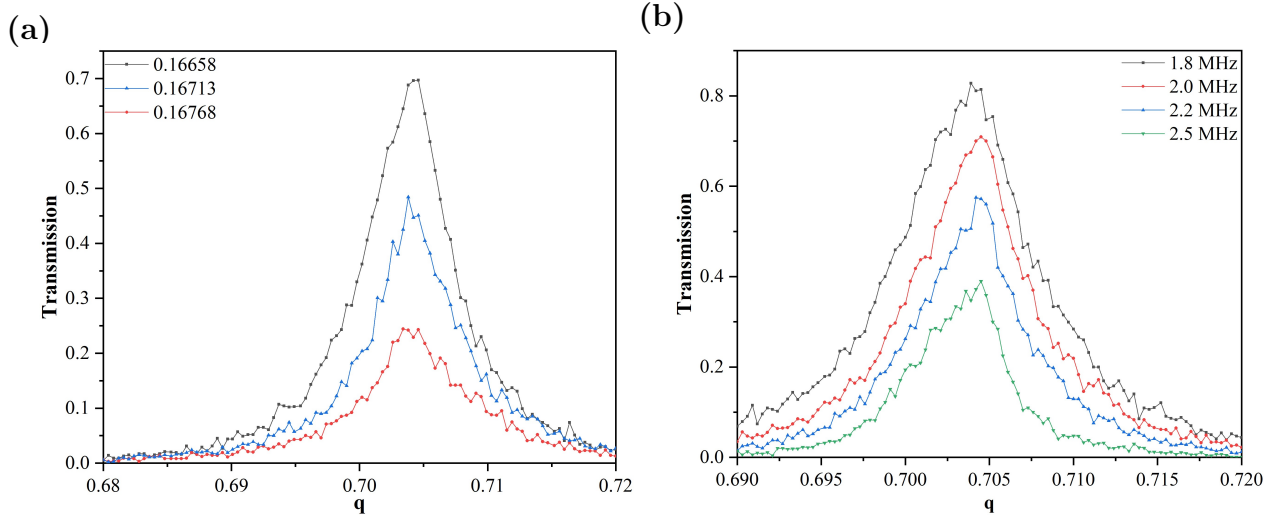

Fig.~\ref{fig2} (a) illustrates the effect of the scan parameter $\lambda$ on the transmission profile of a symmetric QMF at a fixed frequency of $\omega/2\pi = 2$~MHz. Fig.~\ref{fig2}(a) shows that the maximum transmission is reduced and the transmission contour becomes sharper with increasing $\lambda$, consistent with the QMF literature \cite{sysoev2022,douglas2002influence}. The transmission is sensitive beyond three decimal places of $\lambda$, demonstrating the precision of our study with SIMION. 

Fig.~\ref{fig2} (b) shows the effect of varying the driving frequence in the range $1.8$ MHz and $2.5$ MHz while keeping $\lambda$ constant at $0.16658$. The ions experience a larger number of rf cycles at higher frequency, thus reducing the peak and the width of the transmission contour, as consistent with the literature\cite{dawson2013quadrupole,douglas2009}

\begin{figure}[h!]
    \centering
    \includegraphics[width=0.65\linewidth]{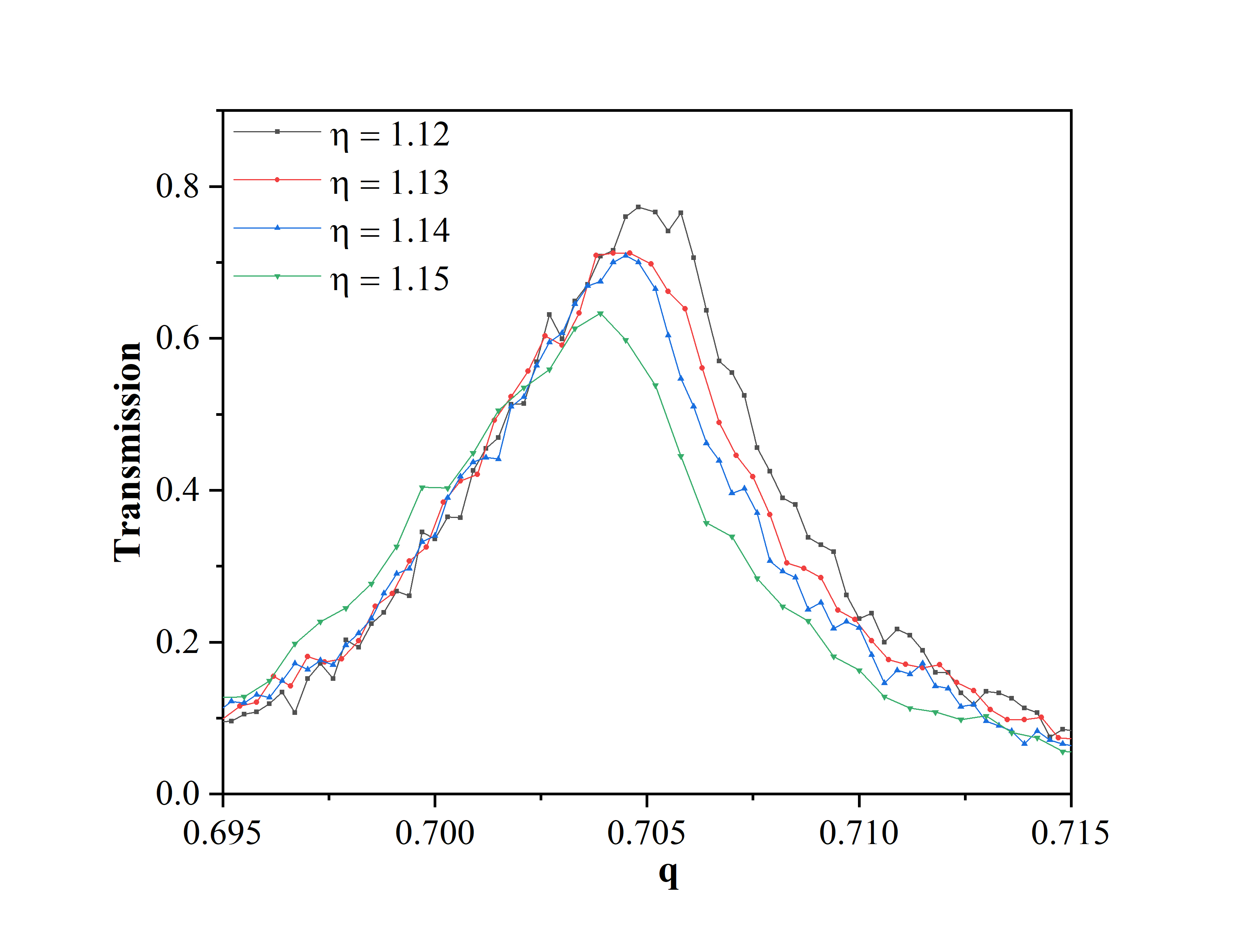}
    \caption{Transmission characteristics of symmetric QMF for different geometric parameter $\eta$. Simulation were performed with $\lambda=0.16658$ and $\omega/2\pi$ = 2 MHz.}
    \label{fig3}
\end{figure}

All subsequent transmission studies have been carried out at a fixed frequency of $\omega/2\pi = 2$~MHz and a scan parameter of $\lambda = 0.16658$. Transmission profiles for symmetric QMFs with $\eta = 1.12$, $1.13$, $1.14$ and $1.15$ are shown in fig.~\ref{fig3}, revealing subtle yet significant variation. The transmission curve for $\eta = 1.14$ is the narrowest, consistent with the expectation of minimal higher-order field distortion at this geometry.

\section{Asymmetric QMF}

Asymmetry in a quadrupole mass filter can arise unintentionally due to mechanical tolerances or machining imperfections~\cite{taylor2008prediction,ding2003quadrupole}. However, deliberate asymmetries can also be introduced into circular-rod QMFs to study or exploit specific field effects. For instance, a diagonally opposite pair of electrodes can be radially displaced~\cite{sysoev2022}, or designed with diameters different from the other pair~\cite{ding2003quadrupole}; alternatively, a single electrode may be fabricated with a diameter different from the remaining three~\cite{mandal2024}. 

In the present study, a controlled asymmetry is introduced by modifying the diameter of one electrode while keeping its position fixed, as illustrated in fig.~\ref{fig1}(c). This configuration distorts the ideal quadrupole field and renders the field radius $r_0$, and consequently the $q$ parameter ill-defined. 
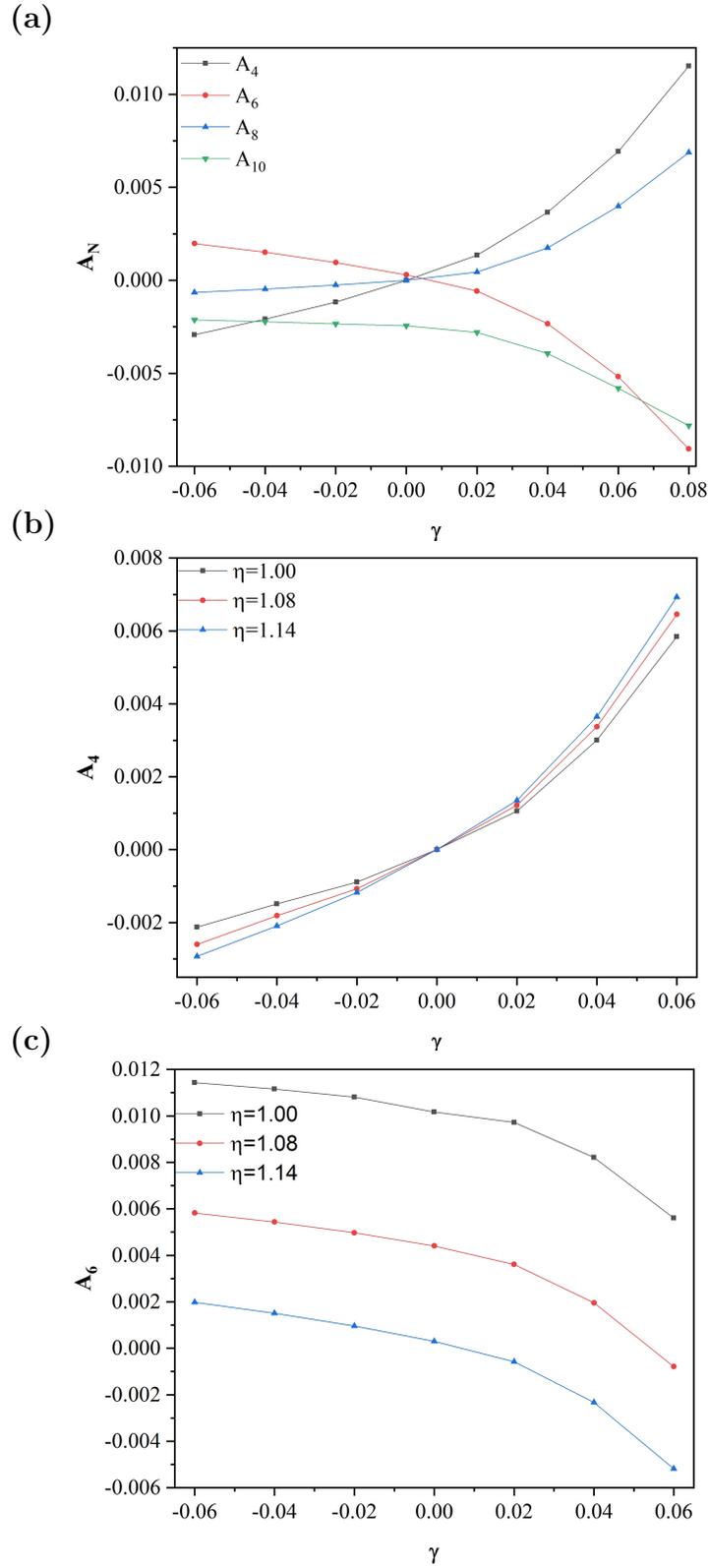
\begin{figure}
    \begin{subfigure}[b]{0.525\textwidth}
    \begin{picture}(0,0)
            \put(-20,200){\textbf{(a)}}
        \end{picture}
        \includegraphics[width=\textwidth]{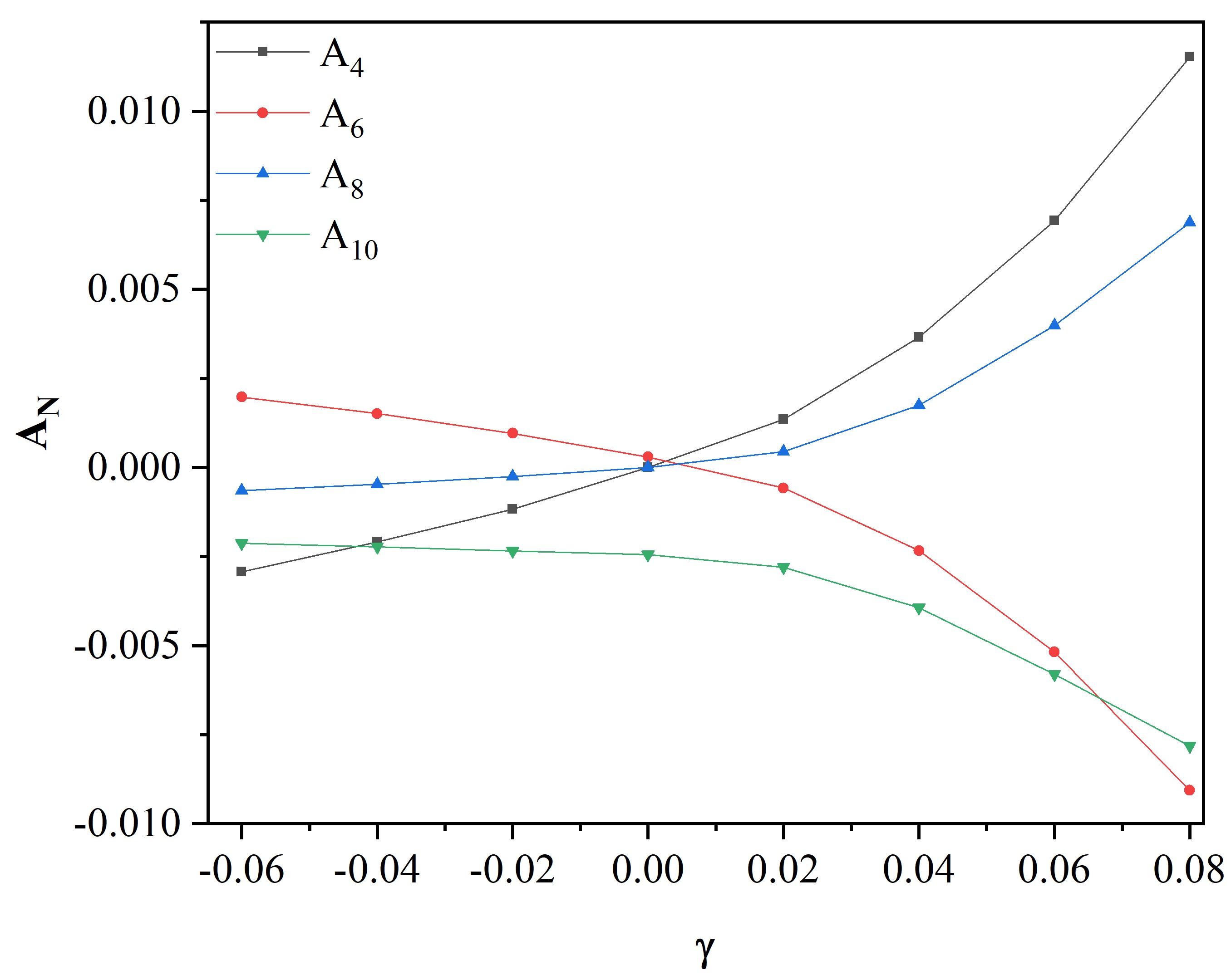 }
    \end{subfigure}
        \par\vspace*{-1.2mm}    
    \begin{subfigure} [b]{0.525\textwidth}
    \begin{picture}(0,0)
            \put(-20,200){\textbf{(b)}}
        \end{picture}
        \includegraphics[width=\textwidth]{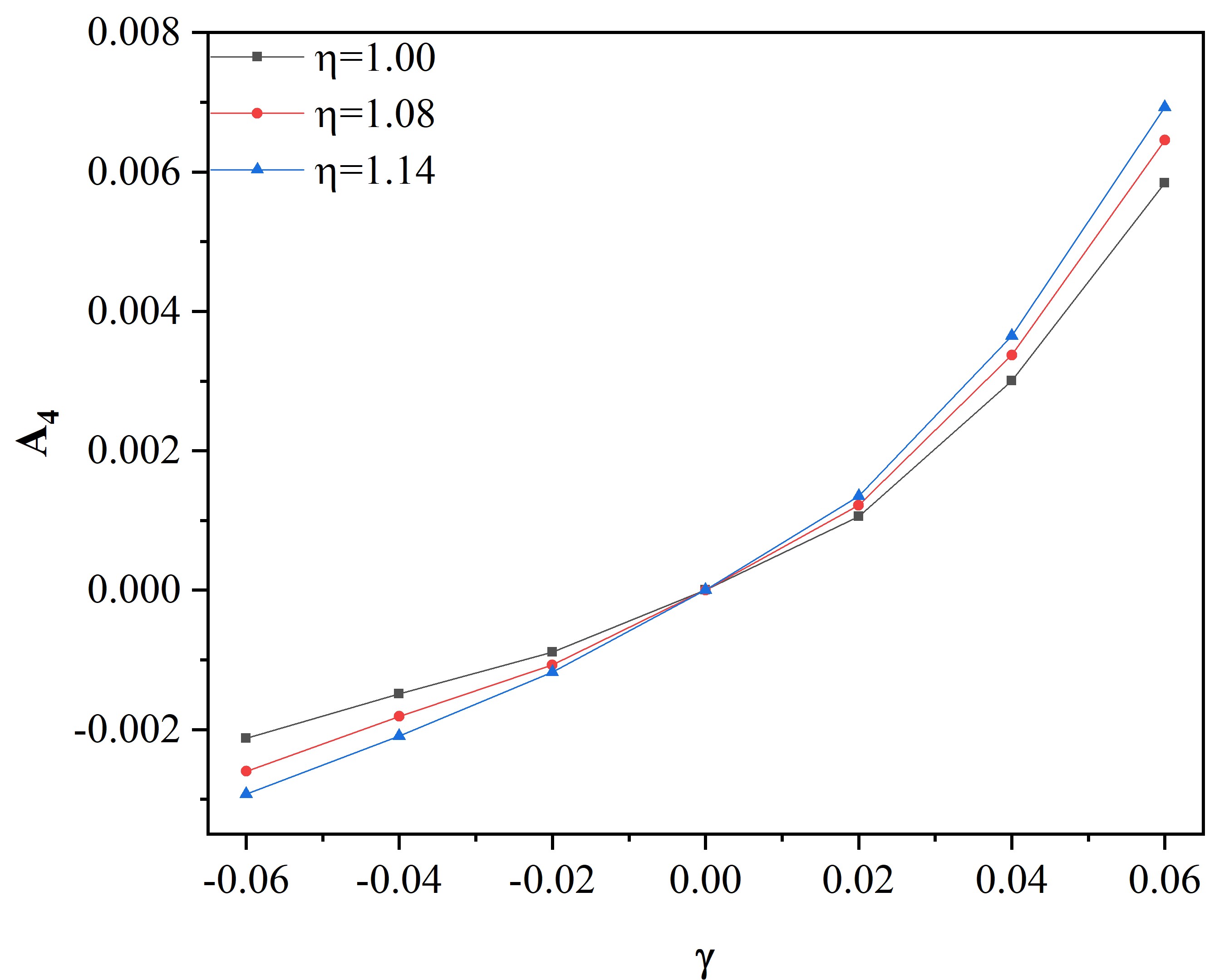} 
    \end{subfigure}
        \par\vspace*{-0.5mm}
    \begin{subfigure}[b]{0.525\textwidth}
    \begin{picture}(0,0)
            \put(-20,200){\textbf{(c)}}
        \end{picture}
        \includegraphics[width=\textwidth]{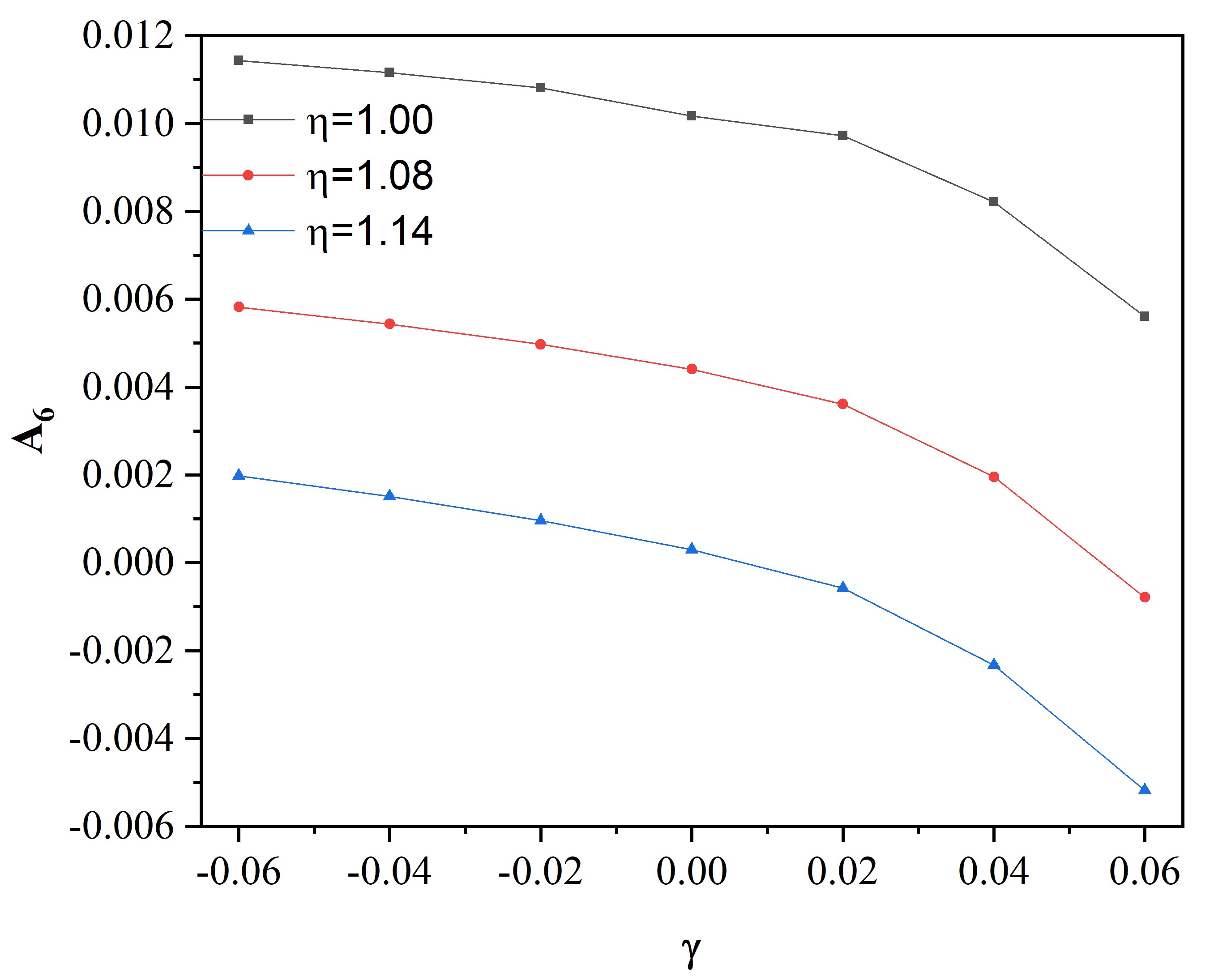} 
    \end{subfigure}
    \caption{(a) Variation of $A_4$, $A_6$, $A_8$ and $A_{10}$ as a function of $\gamma$, for $\eta=1.14$; (b) Variation of $A_4$ as a function of $\gamma$ for $\eta = 1.00$, $1.08$ and $1.14$; (c) Dependence of $A_6$ on $\gamma$ for QMF geometries with $\eta = 1.00$, $1.08$ and $1.14$. The solid lines connecting the data points serve merely as visual guides.} 
    \label{fig4}
\end{figure}

\subsection{Multipole fields}

To investigate the influence of multipole fields and their impact on transmission characteristics in asymmetric QMFs, a fixed field radius of $r_0 = 5.0$~mm is adopted for all configurations. The radius of the three identical electrodes is calculated as $r = \eta r_0$ for a given $\eta$, while the radius of the fourth (asymmetric) electrode, denoted as $r_a$ (fig.~\ref{fig1}(c)), is varied in increments of $0.1$~mm relative to $r$. A dimensionless design parameter $\gamma = (r_a - r)/r_0$ is introduced to quantify the degree of asymmetry. The electric potential within the mass filter is simulated using SIMION, following the same methodology as employed for the symmetric configurations. The resulting potential distributions for each asymmetric design are fitted to eq.~\ref{eq2}, assuming a constant field radius of $r_0 = 5.0$~mm. The amplitudes of the spatial harmonic components, $A_N$, are extracted for $N$ ranging from $0$ to $10$.

\begin{table*}[h!]
\centering \label{table2}
\caption{\label{asym-coeff}Amplitude $A_N$ vs. asymmetric parameter($\gamma$) for an asymmetric setup of $\eta$=1.14}
\begin{tabular}{c|cccccc}
\hline
$\gamma$ & $A_2$ & $A_4 \times 10^3$ & $A_6 \times 10^3$ & $A_8 \times 10^3$ & $A_{10} \times 10^3$ \\  
\hline
-0.06 & 0.971  & -2.928  & 1.978   & -0.655  & -2.128   \\
-0.04 & 0.981  & -2.093  & 1.509   & -0.468  & -2.228   \\
-0.02 & 0.992  & -1.172 & 0.958  & -0.256  & -2.339  \\
\hline
0 & 1.002  &  0 & 0.295   & 0 &-2.443  &   \\
\hline
0.02 & 1.013   & 1.353   & -0.579   & 0.444  & -2.797   \\
0.04 & 1.024   & 3.650   & -2.333   & 1.739   & -3.939   \\
0.06 & 1.034  & 6.928   & -5.178   & 3.980    & -5.803   \\

\hline
\end{tabular}
\label{table2}
\end{table*}

The multipole coefficients $A_N$ extracted for various asymmetric designs with $\eta = 1.14$ are tabulated in table \ref{table2}. Coefficients with magnitudes below $10^{-5}$ are taken as zero, and the corresponding multipole contributions are considered absent. A key observation is the emergence of octupole and hexadecapole components in asymmetric configurations, which are otherwise absent in the symmetric circular-rod geometry. This is consistent with experimental observations of nonlinear resonances attributed to octupole fields in linear ion trap with similar asymmetry~\cite{mandal2024}, as well as previous reports of octupole contributions in QMFs with opposite electrodes of unequal diameters~\cite{ding2003quadrupole}.

A systematic dependence of $A_N$ on the asymmetry parameter $\gamma$ is observed, as illustrated in fig.\ref{fig4}, for $\eta$ = 1.14. Specifically, $A_4$ increases for $\gamma > 0$ and decreases for $\gamma < 0$, whereas $A_6$ exhibits the opposite trend, suggesting a partial compensatory effect between octupole and dodecapole components. A similar complementary behavior is noted between $A_8$ and $A_{10}$, particularly for positive $\gamma$. Overall, both table \ref{table2} and fig.~\ref{fig4} demonstrate that $|A_N|$ increases with increasing asymmetry, regardless of the sign of $\gamma$, with the exception of $|A_{10}|$, which shows minimal variation for $\gamma < 0$. These results confirm that any deviation from symmetry enhances the presence and magnitude of higher-order multipole fields.

The dodecapole component contributes significantly to the field distortion in symmetric QMF geometries with $\eta < 1.14$. For instance, our simulations yield $A_6 = 4.41 \times 10^{-3}$ for $\eta = 1.08$ and $A_6 = 1.02 \times 10^{-2}$ for $\eta = 1.00$. Upon introducing asymmetry to the $\eta = 1.08$ configuration, additional multipole components such as $A_4$ and $A_8$ emerge, exhibiting similar trends with the asymmetry parameter $\gamma$ as observed earlier for $\eta = 1.14$ (fig.~\ref{fig4}). Notably, $A_4$ shows a complimentary  variation with $A_6$. A similar analysis performed for the $\eta = 1.00$ geometry confirms these trends, reinforcing the generality of the observed multipole behavior across symmetric and asymmetric QMF designs.

\subsection{Transmission characteristics}
Higher-order multipole components play a critical role in shaping the transmission profile of a QMF, directly impacting resolution and mass selectivity. In symmetric configurations, the field is predominantly quadrupolar. However, asymmetries, whether introduced intentionally or arising from fabrication imperfections, induce additional multipole components, most notably the octupole harmonics. To systematically probe these effects, a detailed SIMION-based transmission studies are conducted across a range of asymmetric QMF geometries, focusing on the influence of the higher order terms.

\begin{figure}%
         
    \begin{subfigure}[b]{0.9\textwidth}
    \begin{picture}(0,0)
            \put(-20,200){\textbf{(a)}}
        \end{picture}
        \includegraphics[width=\textwidth]{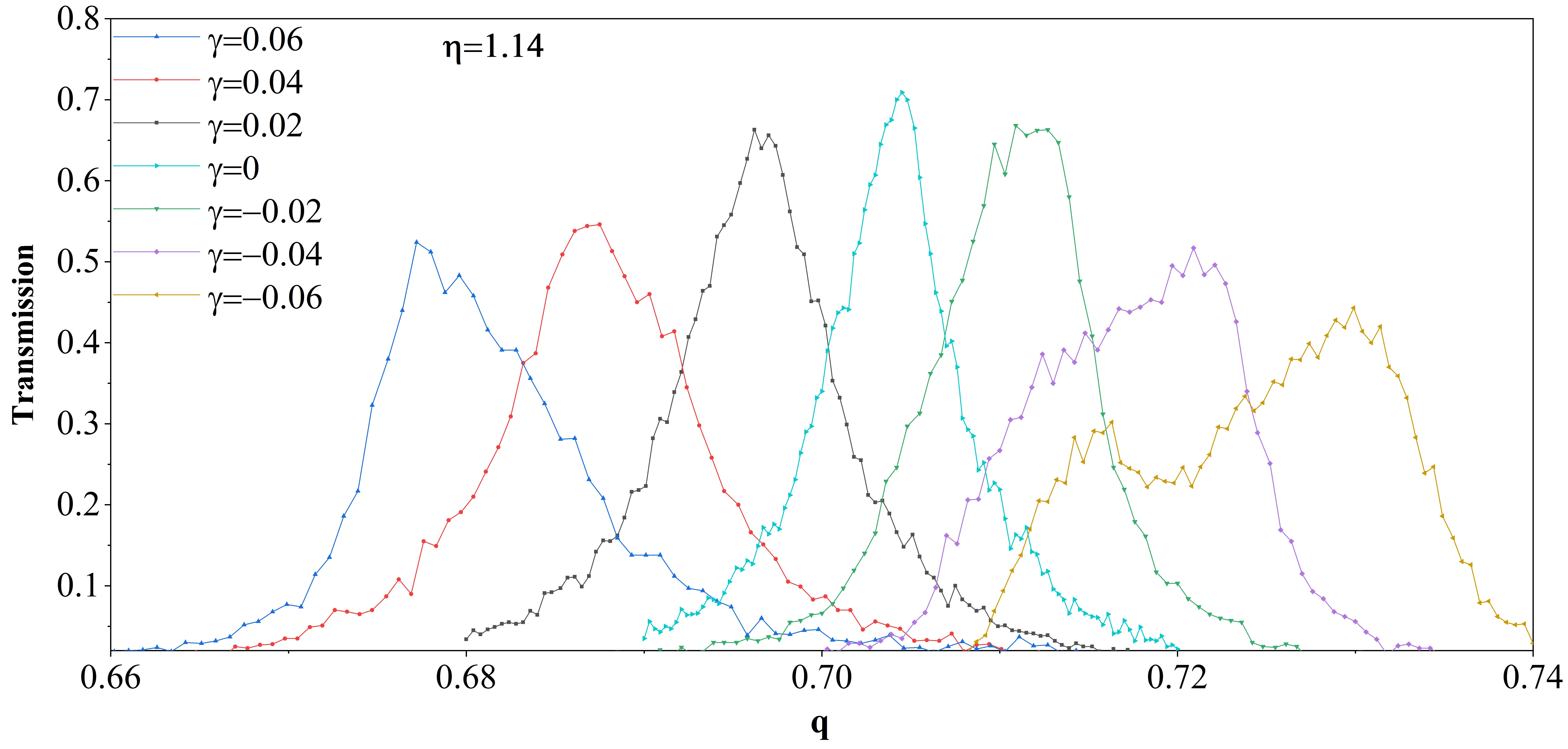}
    \end{subfigure}
        \par\vspace*{-1.2mm}    
    \begin{subfigure} [b]{0.9\textwidth}
    \begin{picture}(0,0)
            \put(-20,200){\textbf{(b)}}
        \end{picture}
        \includegraphics[width=\textwidth]{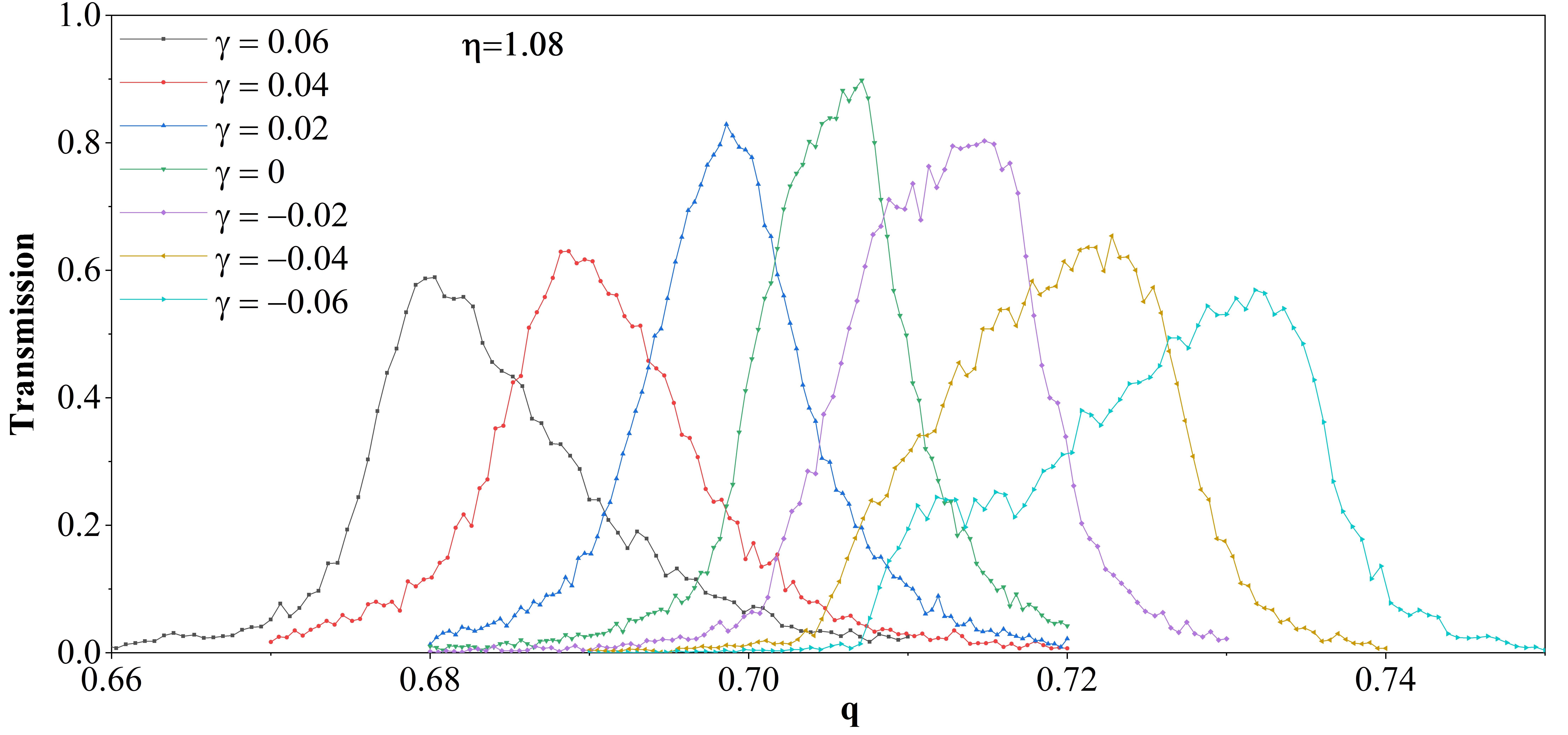} 
    \end{subfigure}
        \par\vspace*{-0.5mm}
    \begin{subfigure}[b]{0.9\textwidth}
    \begin{picture}(0,0)
            \put(-20,200){\textbf{(c)}}
        \end{picture}
        \includegraphics[width=\textwidth]{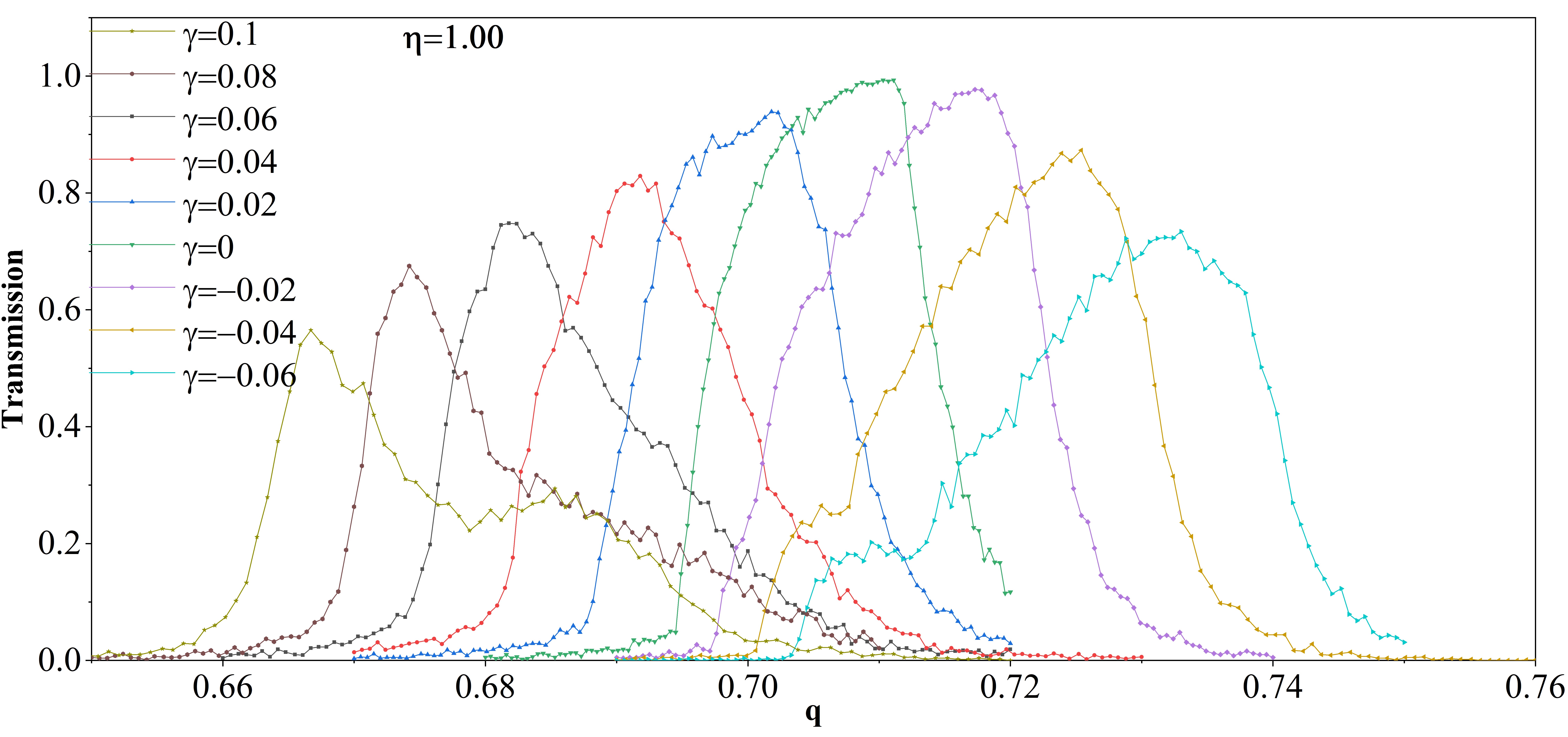} 
    \end{subfigure}
    \caption{The transmission contours for different asymmetric designs added to (a) $\eta=1.14$ (b) $\eta=1.08$ and (c) $\eta=1.00$. The asymmetric parameters are mentioned in the legends.} 
    \label{fig5}
\end{figure}
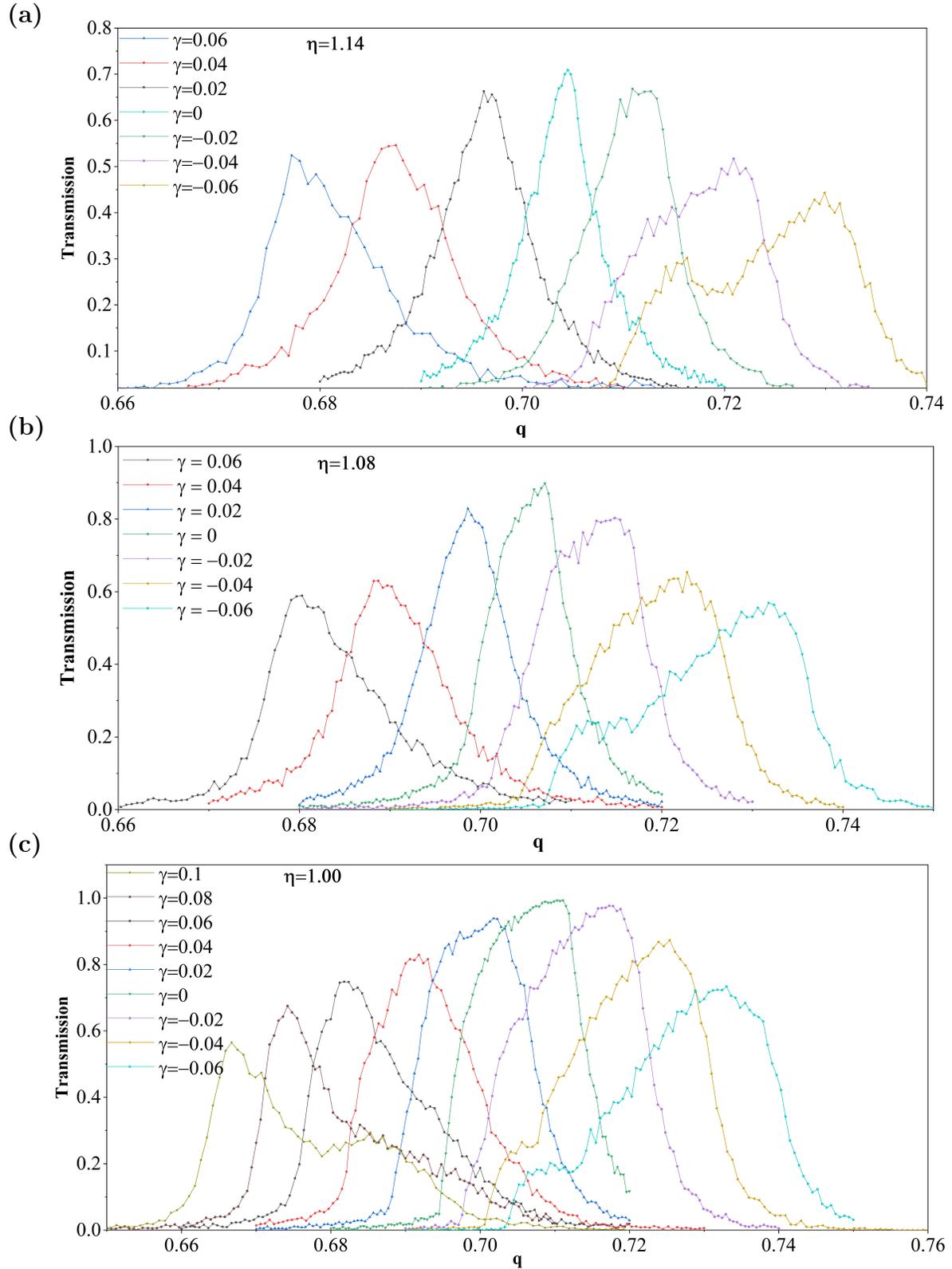

For the symmetric configuration at $\eta = 1.14$, the transmission curve exhibits a sharp peak centered at $q = 0.7045$, as expected for the scan parameter $\lambda = 0.16658$. This peak corresponds to maximum transmission through the filter. As asymmetry is introduced by modifying the radius of the fourth electrode (quantified by the asymmetry parameter $\gamma$), the position of the transmission peak shifts: for $\gamma > 0$ (where radius of the fourth electrode is larger than the other three), the peak shifts to lower $q$ values, whereas for $\gamma < 0$ (in which the fourth electrode is designed with a smaller radius than the other three), it moves to higher $q$. The transmission profile as a function of $q$ has been shown in fig.~\ref{fig5}(a) for $\eta$ = 1.14. The maximum transmission falls and the transmission contour gets broadened as the asymmetry increases on either side. 
Of particular note is the emergence of a secondary transmission feature or precursor peak at $\gamma = -0.6$, as reported earlier with an imperfect quadrupole mass spectrometer.~\cite{taylor2008prediction}

To verify the generality of these observations, similar analyses were performed for asymmetric variations based on symmetric QMF structures with $\eta = 1.08$ and $\eta = 1.00$ and illustrated in fig.~\ref{fig5}(b) and fig.~\ref{fig5}(c) respectively. In both cases, the transmission trends are similar to those seen for $\eta = 1.14$: asymmetric perturbations led to a systematic shift and degradation of the transmission profile, confirming that the observed behavior is robust across different base geometries. These results reinforce the conclusion that even modest deviations from symmetry can significantly alter the ion transmission characteristics, highlighting the necessity for precise mechanical tolerances in QMF design and fabrication.

\subsection{Resolution}

The resolution of the mass filter, $R$, is influenced by multiple factors, including the ratio $U/V_0$ (or equivalently, the scan parameter $\lambda$), the number of rf cycles an ion experiences and the geometry of the electrodes~\cite{douglas2009}. Here we have defined $R=q/\Delta q$ to compare the performance of various asymmetric designs, where $\Delta q$ describes the full width at half-maxima (FWHM) of the transmission contour. A narrower and sharper transmission profile indicates higher resolution. It is evident from fig.~\ref{fig5} that as the magnitude of asymmetry $|\gamma|$ increases, the height of the transmission peak decreases, and the profile broadens. This broadening thus has a direct impact on $R$.

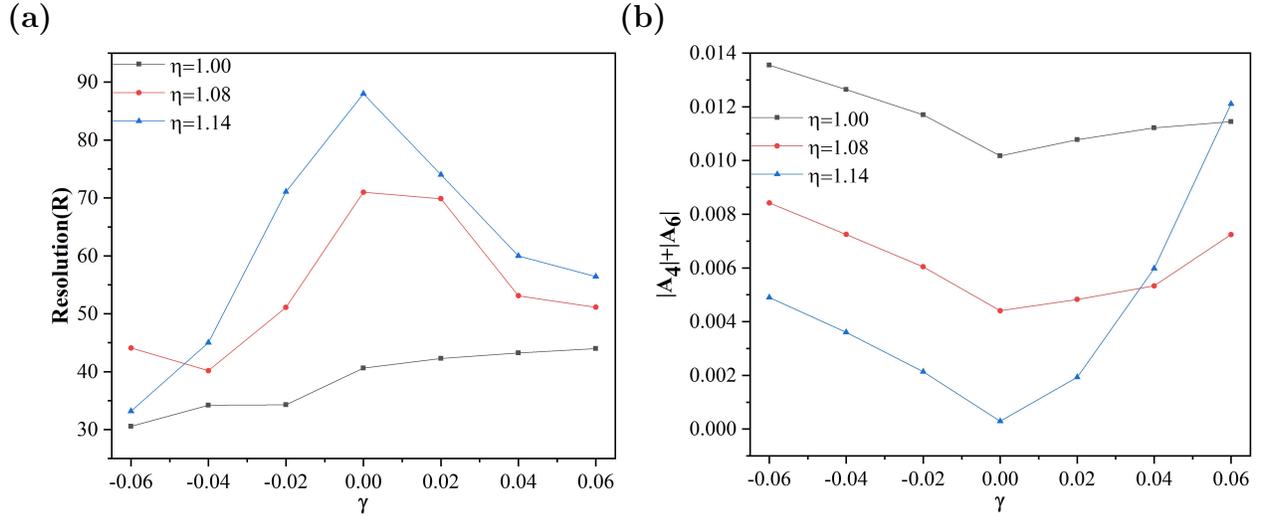
\begin{figure}[H]
    \centering
    \begin{subfigure}[b]{0.47\textwidth}
        \begin{picture}(0,0)
            \put(-10,185){\textbf{(a)}}
        \end{picture}
        \includegraphics[width=\textwidth]{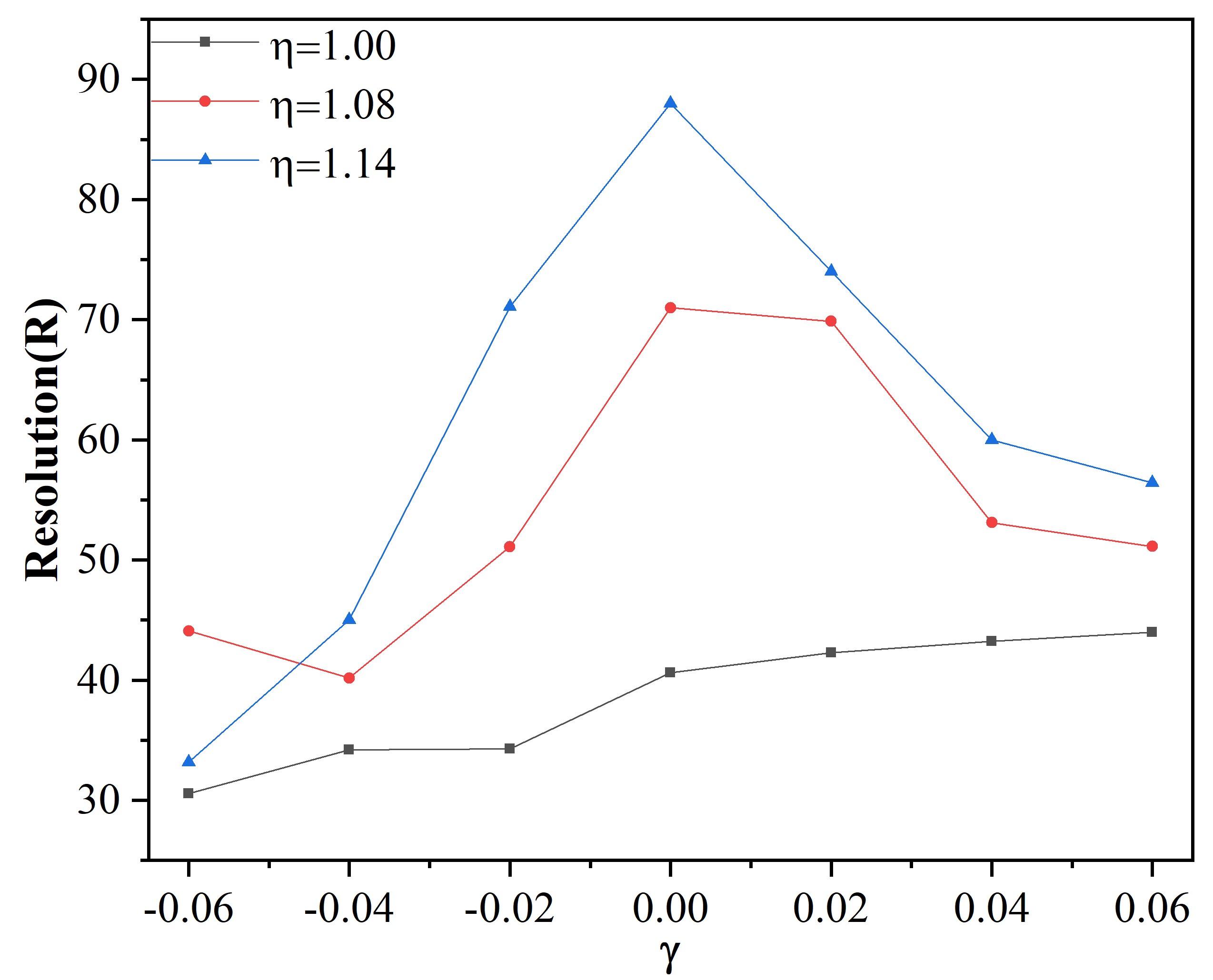}
        \label{fig:a}
    \end{subfigure}
    \hspace{1mm} 
    \begin{subfigure}[b]{0.49\textwidth}
        \begin{picture}(0,0)
            \put(-10,185){\textbf{(b)}}
        \end{picture}
        \includegraphics[width=\textwidth]{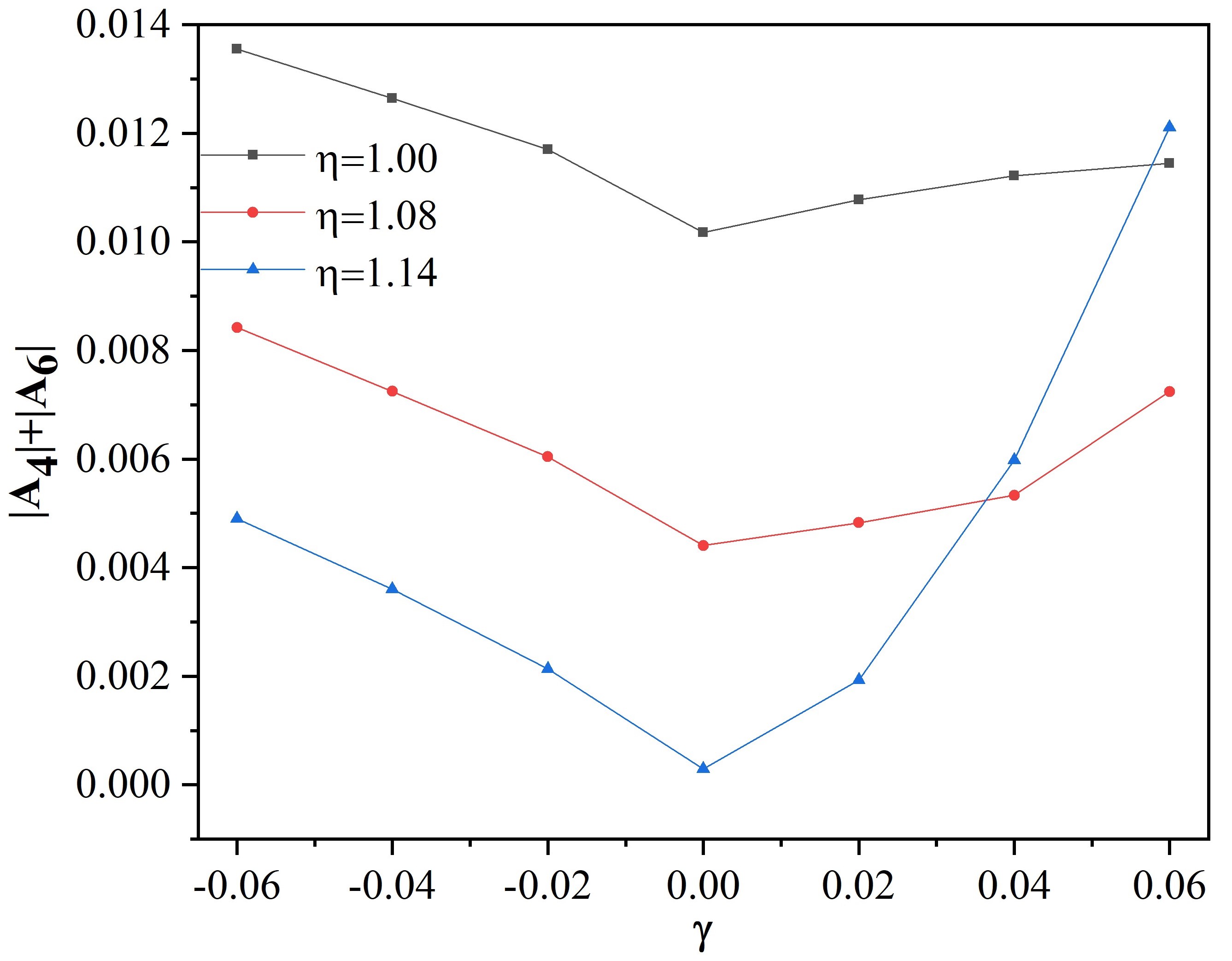}
        \label{fig:b}
    \end{subfigure}
    \caption{(a) Resolution parameter $R$ with the asymmetry parameter for different $\eta$; (b) $|A_4|+|A_6|$ describing the combined weightage of the octupole and dodecapole fields; The solid lines connecting the data points serve merely as visual guides.}
    \label{fig6}
\end{figure}

From fig.~\ref{fig5}, the calculated resolutions for the symmetric QMF configurations at $\lambda = 0.16658$ are $R = 88.1$ for $\eta = 1.14$, $71.0$ for $\eta = 1.08$, and $40.8$ for $\eta = 1.00$. These values confirm that resolution degrades with increasing higher-order field contributions. Fig.~\ref{fig6}(a) plots $R$ as a function of the asymmetry parameter $\gamma$ for different configurations, showing a systematic decline in resolution with increasing asymmetry. In particular, for each geometry, the maximum resolution is obtained in the symmetric case, with $\eta = 1.14$ providing the best overall performance. 

With increasing radius of the odd electrode ($\gamma > 0$), the octupole coefficient $A_4$ becomes positive while the dodecapole coefficient $A_6$ decreases below zero for $\eta = 1.14$, as shown in fig.~\ref{fig4}(b) and fig.~\ref{fig4}(c). Although a partial compensation between the octupole and dodecapole contributions might be expected, the balance is incomplete, since $A_4 + A_6 \neq 0$. Additionally, the scaling behavior of these multipoles differs: the octupole potential scales as $1/r_0^4$, whereas the dodecapole component scales as $1/r_0^6$. The resulting distortion in the quadrupole field leads to a degradation in resolution for $\gamma > 0$.

Conversely, for negative $\gamma$, where the radius of the odd electrode is reduced, $A_4$ becomes negative and $A_6$ increases above zero. However, $|A_4| > |A_6|$ in this case for $\eta = 1.14$, indicating that the imbalance persists and resolution continues to deteriorate. In asymmetric QMF designs with $\eta = 1.08$, a similar imbalance is observed ($|A_4 + A_6| \neq 0$), and the transmission characteristics replicate those seen for $\eta = 1.14$.

A mild exception arises for $\eta = 1.00$ with $\gamma > 0$, where a strong dodecapole component, even in the symmetric configuration, makes the transmission contour quite broad and resolution becomes poor to proceed for a conclusive study with such quantitative approach.

Fig.~\ref{fig6}(b) illustrates the variation of $|A_4| + |A_6|$ with $\gamma$ at $\eta$ corresponding to 1.00, 1.08 and 1.14, highlighting the cumulative distortion arising from these dominant multipoles. It is worthwhile to mention that the hexadecapole and icosapole potential components scale as $1/r_0^8$ and $1/r_0^{10}$ respectively with insignificant effect on transmission, particularly on a beam radius $0.1r_0$, and hence are not considered in the performance analysis.

\section{Conclusion}

The multipole fields and the performance of both symmetric and asymmetric QMF have been systematically investigated, with particular focus on the influence of key parameters: the geometry factor ($\eta$), mass scan parameter ($\lambda$), driving frequency ($\omega$), and the asymmetry parameter ($\gamma$). For the symmetric QMF, simulations within the resolution limits of SIMION indicate that an optimal geometry at $\eta = 1.14$ minimizes quadrupole field distortion and yields maximum mass resolution, which exhibits sensitivity to the parameter $\lambda$, consistent with literature.

Linear QMF with an asymmetric circular rod appears with octupole and hexadecapole fields in addition to dodecapole and icosapole fields that alter the ion transmission characteristics and affect the resolution.  Correlations have been identified between the multipole coefficients $A_4$ and $A_6$, as well as $A_8$ and $A_{10}$, across varying $\eta$, as a function of the asymmetry parameter $\gamma$.

Transmission analysis reveal that for a fixed $\eta$, the highest resolution is consistently achieved under symmetric electrode configurations. Even slight deviations from symmetry, such as machining imperfections in a single electrode, result in measurable losses in resolution, despite compensatory conditions like $A_4 + A_6 = 0$. This underscores the limited utility of attempting to balance one multipole component with another. Notably, the resolution parameter $R$ exhibits a strong empirical correlation with the combined multipole strength $|A_4| + |A_6|$, a relationship that may stem from alterations in the stability diagram due to octupole and dodecapole perturbations. This finding warrants further theoretical exploration. It also demonstrates how intentional octupole perturbations can be introduced into linear ion traps, with potential applications in the controlled study of trapped ion dynamics.

\begin{acknowledgement}

ND thanks SERB/ANRF India (CRG/2023/001529) and BRNS India (58/14/21/2023-BRNS12329) for funding. SJ acknowledges Indian Association for the Cultivation of Science India for research fellowship.

\end{acknowledgement}


\bibliography{reference}

\end{document}